\documentclass{article}
\usepackage{spconf,amsmath,graphicx}
\usepackage{booktabs}
\usepackage{enumitem}
\usepackage{dirtytalk}
\setlist{nosep, leftmargin=14pt}

\usepackage{mwe} 

\title{Utilizing Radiomic Feature Analysis for Automated MRI Keypoint Detection: Enhancing Graph Applications}
\name{\begin{tabular}{c}
   Sahar Almahfouz Nasser, Shashwat Pathak, Keshav Singhal, Mohit Meena, Nihar Gupte, Ananya Chinmaya,\\
   Prateek Garg, and Amit Sethi\\
  \end{tabular}\thanks{}}

\address{Indian Institute of Technology Bombay  }
\begin{document}
%

\maketitle

\begin{abstract}

Graph neural networks (GNNs) present a promising alternative to CNNs and transformers in certain image processing applications due to their parameter-efficiency in modeling spatial relationships. Currently, a major area of research involves the converting non-graph input data for GNN-based models, notably in scenarios where the data originates from images. One approach involves converting images into nodes by identifying significant keypoints within them. SuperRetina, a semi-supervised technique, has been utilized for detecting keypoints in retinal images. However, its limitations lie in the dependency on a small initial set of ground truth keypoints, which is progressively expanded to detect more keypoints. Having encountered difficulties in detecting consistent initial keypoints in brain images using SIFT and LoFTR, we proposed a new approach: radiomic feature-based keypoint detection. Demonstrating the anatomical significance of the detected keypoints was achieved by showcasing their efficacy in improving registration processes guided by these keypoints. Subsequently, these keypoints were employed as the ground truth for the keypoint detection method (LK-SuperRetina). Furthermore, the study showcases the application of GNNs in image matching, highlighting their superior performance in terms of both the number of good matches and confidence scores. This research sets the stage for expanding GNN applications into various other applications, including but not limited to image classification, segmentation, and registration.
\end{abstract}
\section{Introduction}
\label{sec:intro}

Graph neural networks (GNNs) have shown promising results for reducing the computational requirements for certain image processing tasks as shown in~\cite{he2023generalization}. However, converting images into graphs is an active area of research. Many papers have tried breaking down images into patches and treating each patch as a node in a graph~\cite{he2023generalization}. Our new method relies on detecting important keypoints in the images along with their features and making graphs from them.

Detecting important keypoints in certain types of images, such as magnetic resonance images of the brain is not straightforward due to the lack of well-defined landmarks. In various registration competitions, such as the BraTSReg challenge~\cite{baheti2021brain}, experts have had to mark specific landmark points on 3D brain images to help the participants align images and evaluate how well their registration algorithms worked. But this marking is neither easy nor fast. Also, only a few points, around 6 to 50 per set of images, get marked.



Traditional keypoint detection algorithms such as SIFT~\cite{lindeberg2012scale} fall behind deep learning-based keypoint detection algorithms which have different types include: supervised, unsupervised, and semi-supervised methods. Some examples are UnsuperPoint \cite{christiansen2019unsuperpoint}, SuperPoint \cite{detone2018superpoint}, GLAMpoints \cite{truong2019glampoints}, and SuperRetina \cite{liu2022semi}.

After trying and failing to find reliable keypoints in brain images using methods such as SIFT~\cite{lindeberg2012scale} and LoFTR~\cite{sun2021loftr}, we proposed a new algorithm. Our proposed method finds keypoints using radiomic features in brain images and their feature vectors. We proved these keypoints are important landmarks in brain images by showing how they help in registering brain images. Then we used this dataset we made, which consists of MRI images from OASIS dataset~\cite{marcus2007open} and the keypoints we detected, to train the LK-SuperRetina algorithm~\cite{almahfouz2023reverse} to detect new keypoints. We showed that using a GNN like SuperGlue~\cite{sarlin2020superglue} for image matching improves the features of detected keypoints. This method of detecting keypoints opens doors for using graph-based neural networks for different tasks like brain classification, segmentation, and registration.

\section{Proposed Method}
\label{sec:proposed_method}

We introduced an approach to keypoint detection based on radiomic features. Given an image alongside its segmentation labels, our method identifies radiomic keypoints as the centers of radiomic segmentation labels within the image. These radiomic features encompass a range of intensity and shape characteristics specific to the regions defined by the segmentation map. To compute these features, we utilized the Pyradiomic library~\cite{van2017computational}.

\begin{figure*}
  \centering
  \includegraphics[width=2.0\columnwidth]{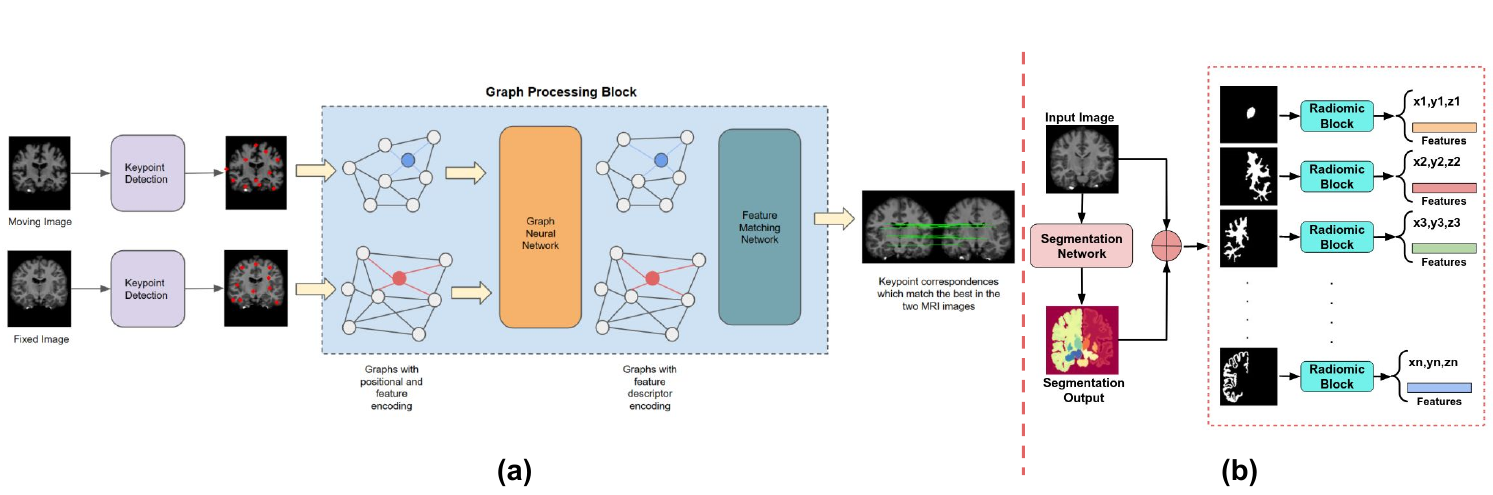}
  \caption{(a) The image matching pipeline, encompassing keypoint detection using neural networks, graph formation from detected keypoints, graph neural network (GNN) processing to enhance keypoint features, and a dedicated head for keypoint matching. (b) Radiomic features-based Keypoint detection method.}
  \label{pipeline}
\end{figure*}

Radiomic keypoints are closely tied to segmentation regions predicted by neural networks. We trained Swin UNetR~\cite{hatamizadeh2021swin} to predict the segmentation maps of brain images and used them as masks to extract the keypoints. Each segmentation mask yields a keypoint location, accompanied by 53 descriptive radiomic features, see figure ~\ref{pipeline} for more details.
Radiomic keypoints exhibit repeatability across various brain samples even in the presence of varying intensity and non-rigid deformations which make our proposed method a more robust alternative to methods like SIFT~\cite{lindeberg2012scale}. 
These detected keypoints can serve as a initial keypoints for detecting additional keypoints through deep learning techniques as we will show in our results, in which we trained LK-SuperRetina~\cite{almahfouz2023reverse} to automatically detect these ground truth keypoints and extra keypoints. This underscores the significance of these keypoints as essential landmarks within brain images.




\section{Experiments and Results}

This section begins with an introduction to the dataset used in all our experiments. Following that, we explore various approaches applied for keypoint detection, explaining how the detected keypoints were utilized to improve registration and identify additional keypoints. Finally, we will conclude with our results on GNN-based image matching.

\subsection{OASIS dataset}
\label{dataset:oasis}

The OASIS dataset, as described in~\cite{marcus2007open}, contains MRI data obtained from 414 subjects. This dataset has been divided into three separate subsets for training, validation, and testing, following a ratio of 314:50:50, respectively. Each subject in the dataset has T1-weighted scan, as well as segmentation masks of various regions of the brain. This dataset incorporates three distinct types of brain segmentation: a four-label mask, a thirty-five label mask, and a twenty-four label mask.The 3D T1-weighted scans and their corresponding masks are of resolution (160, 192, 224). Additionally, the 2D T1-weighted scans and their corresponding masks have a resolution of (160, 192).

\subsection{SIFT}

SIFT~\cite{lindeberg2012scale} was initially chosen for experimentation due to its reputation in keypoint detection. Utilizing gradients, SIFT excels in identifying scale and rotation invariant keypoints. The method calculates orientation gradients across scales, creating a distinctive 128-dimensional vector for each keypoint. 

\begin{figure}
  \centering
  \includegraphics[width=1.0\columnwidth]{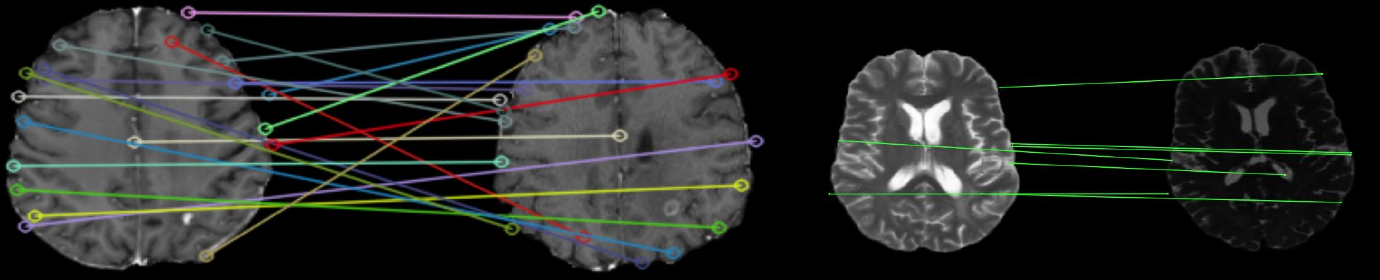}
  \caption{SIFT~\cite{lindeberg2012scale}  and LoFTR~\cite{sun2021loftr} performances in keypoint detection. The left pair of images depicts results for SIFT, while the right one illustrates the performance of LoFTR. It is evident that LoFTR outperforms SIFT in keypoint detection and matching. Nevertheless, LoFTR encounters challenges when the source and target images exhibit varying intensity distributions.}
  \label{sift_lofter}
\end{figure}

 Figure~\ref{sift_lofter} shows  significant issues we encountered  with keypoint detection using SIFT. Notably, there was inconsistency in keypoint locations across different MRI slices within the brain, reflecting a lack of repeatability. Our experiments highlighted SIFT's ineffectiveness under conditions involving large deformations.

\subsection{LoFTR}

Due to SIFT's inability to establish correspondences, we opted to explore a deep learning method namely LoFTR~\cite{sun2021loftr}, which stands for detector-free local feature matching transformer.

In the LoFTR approach, a convolutional neural network (CNN) with an encoder-decoder architecture extracts both low-resolution and high-level features. The LoFTR module, incorporating self-attention and cross-attention blocks, transforms these features, and a differential matching layer offers two methods: optimal transport and dual softmax~\cite{bridle1989training} .
Our implementation used a pretrained model from the Aachen Day Night Dataset~\cite{sattler2018benchmarking}. LoFTR succeeded in generating keypoint matches for similar intensity profiles but faced challenges in other cases similar to SIFT.
These findings emphasize the need for keypoints exhibiting consistency, accurate matching, and intensity profile invariance across the dataset.

\subsection{Radiomic features-based keypoint detection}

Having faced challenges with keypoint detection using SIFT and LoFTR, which proved sensitive to intensity variations, we shifted to our proposed radiomic features-based method. In this section, we prove the importance of the detected keypoints by showing our results on registering brain images among different subjects. we also showcase the applications of the dataset containing the original OASIS scans and the corresponding detected radiomic keypoints on automatic keypoint detection. And finally train a GNN-based image matcher on the same dataset.

\subsubsection{Image registration}

To assess the significance of radiomics keypoints as landmarks, we integrated them into the loss function of a registration network (TransMorph~\cite{chen2022transmorph}). Our findings revealed that incorporating the keypoints' loss led to a notable 3\% enhancement in the registration performance. 

Vision transformers, excelling in capturing long-range spatial relationships, prove effective in medical image tasks due to their large receptive fields. TransMorph~\cite{chen2022transmorph}, a hybrid Transformer-ConvNet model, utilizes these advantages for volumetric medical image registration.
The encoder divides input volumes into 3D patches, projecting them to feature representations through linear layers. Sequential patch merging and Swin Transformer blocks follow. The decoder, with upsampling and convolutional layers, connects to the encoder stages via skip connections, producing the deformation field. 
We contributed by designing a customized loss function for keypoints, utilizing Gaussian-blurred keypoints to create a ground truth heatmap. Combining Dice and inverted Dice losses addressed imbalanced masks, resulting in a \textbf{3\%} Dice score improvement over the OASIS test dataset. TransMorph achieved a dice score of \textbf{0.89} with keypoint loss, compared to 0.86 without, underscoring keypoints' role in enhancing registration performance. A potential avenue for future research involves developing a loss function that considers both the feature descriptors of keypoints and the disparity between the locations of registered keypoints and their counterparts in the target image.

\subsubsection{Automated keypoint detection}

SuperRetina, introduced in~\cite{liu2022semi} is an apdaptive version of SuperPoint model~\cite{detone2018superpoint} for identifying important keypoints in retinal images. Utilizing a semi-supervised learning framework, SuperRetina maximizes the utility of limited labeled retinal image data by combining both supervised and unsupervised techniques.
Yet, its utilization requires an initial set of ground truth keypoints to initiate the process, subsequently increasing the detected keypoints iteratively. In our approach, we use our radiomic keypoints as the initial sets for OASIS images.

LK-SuperRetina~\cite{almahfouz2023reverse} which is a modified version of SuperRetina consists of an encoder for downsampling, along with two decoders—one for keypoint detection and another for descriptor generation. Keypoint detection utilizes a mix of labeled and unlabeled data, while descriptor training employs self-supervised learning.

Following the U-Net~\cite{ronneberger2015u} design, LK-SuperRetina's shallow encoder begins with a single convolutional layer, followed by three blocks containing two convolutional layers, a $2 \times 2$ max-pooling layer, and ReLU activation. The keypoint decoder has three blocks with two convolutional layers, ReLU activation, and concatenation block. The detection map ($P$) is generated through a convolutional block with three convolutional layers and a sigmoid activation.

The loss function combines the detector and the descriptor losses as shown in Equation \ref{det_loss}.

\begin{equation}\label{det_loss}
	l_{det} = l_{clf} + l_{geo}
\end{equation}

The classification loss component ($l_{clf}$) is defined in Equation \ref{detector_loss_part1}, where $\tilde{Y}$ represents the smoothed version of the binary ground truth labels $Y$ of the keypoints after blurring them with a 2D Gaussian.

\begin{equation}\label{detector_loss_part1}
	l_{clf}(I;Y) = 1-\frac{2. \sum_{i,j}(P\circ \tilde{Y})_{i,j}}{\sum_{i,j}(P \circ P)_{i,j}+\sum_{i,j}(\tilde{Y}\circ \tilde{Y})_{i,j}}
\end{equation}

When feeding both the image $I$ and its augmented version $I'$ to the network, two tensors for the descriptors $D$ and $D'$ are obtained. For each keypoint $(i,j)$ in the non-maximum suppressed keypoint set $\tilde{P}$, two distances are computed: $\Phi_{i,j}^{rand}$ between the descriptors of $(i,j)$ in the set $\tilde{P}$ and a random point from the registered heatmap $H(\tilde{P})$, and $\Phi_{i,j}^{hard}$ representing the minimal distance, as depicted in Equation \ref{desc_loss_part1}.

\begin{equation}\label{desc_loss_part1}
	l_{des}(I,H) = \sum_{(i,j)\in \tilde{P}}max(0,m+\Phi_{i,j}-\frac{1}{2}(\Phi_{i,j}^{rand}+\Phi_{i,j}^{hard}))   
\end{equation}

\begin{figure}
  \centering
  \includegraphics[width=0.8\columnwidth]{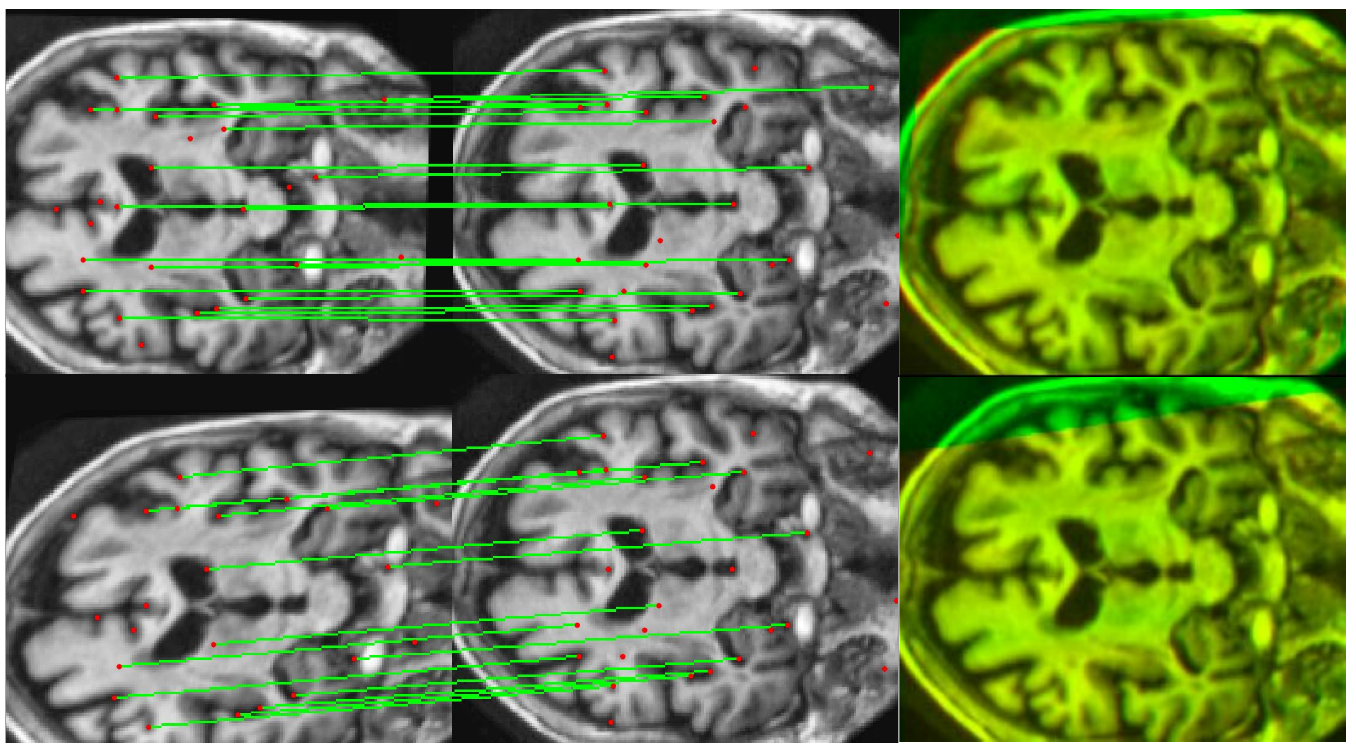}
  \caption{Two examples show the robustness of the keypoint detection model. In both rows, the sequence of images, from left to right, includes: the reference image, the target image, and the registration output. The good matches contribute to aligning the images effectively.}
  \label{keypoints}
\end{figure}

For more in-depth information on the loss function, please refer to the SuperRetina paper~\cite{liu2022semi}.
Figure~\ref{keypoints} shows the results obtained from LK-SuperRetina. As demonstrated, the number of additionally identified keypoints meets expectations, showcasing the network's proficiency in capturing good new keypoints. The network successfully detects both the ground truth keypoints and extra keypoints during the testing phase.

Figure~\ref{keypoints} presents two instances demonstrating the resilience of the keypoint detection model against deformations. The images were deformed randomly by an affine deformation. We passed the original image (target) and the deformed image (reference) seperately to LK-SuperRetina. The model successfully identified corresponding keypoints in both images, as indicated by the number of good matches. We adjusted the thresholds of LK-SuperRetina to detect a smaller set of keypoints for the clarity of the visualization, but in practice, the model can detect over 300 good keypoints.

\subsubsection{Image matching}

Following the detection of keypoints within the brain images, we proceed to construct graphs to be used as inputs of the GNN, for accomplishing specific tasks such as matching in our study. Within this paper, we demonstrate our success in training a GNN-based matcher (SuperGlue~\cite{sarlin2020superglue}) using the graphs formed from the detected keypoints.

SuperGlue designed for matching two sets of local features by identifying correspondences and filtering non-matchable points. Using attention-based graph neural networks, it integrates context aggregation, matching, and filtering within a unified architecture. SuperGlue employs self-attention to enhance the receptive field of local descriptors and cross-attention for cross-image communication. The network handles partial assignments and occluded points by solving an optimal transport problem. With superior performance over other learned approaches, SuperGlue achieves state-of-the-art results in pose estimation for challenging real-world indoor and outdoor environments.

\begin{figure}
  \centering
  \includegraphics[width=1.0\columnwidth]{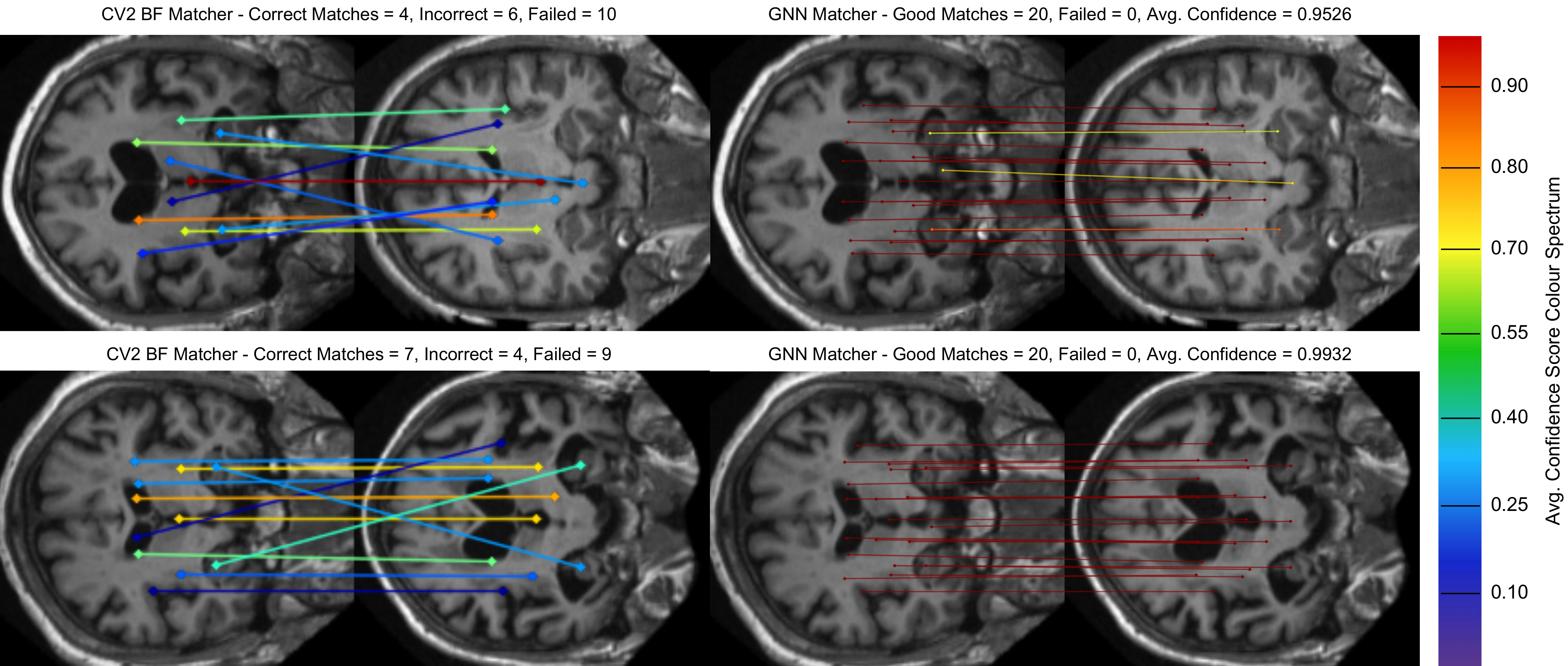}
  \caption{A comparison of the SuperGlue and brute force matcher performance in matching detected keypoints on brain images.}
  \label{matches}
\end{figure}

Figure~\ref{matches} and Table~\ref{tab:results} show a performance comparison between the brute force matcher and SuperGlue across the test dataset. SuperGlue enhances the features of the detected keypoints which improves the matching performance. The identification of keypoints and subsequent graph creation broadens the application of GNNs to various tasks in brain image analysis, extending beyond image matching.

\begin{table}
  \centering
  \begin{tabular}{@{}lr@{}}
    \toprule
    Method &  Avg. No. Good Matches\quad \quad  Confidence Score \\
    \midrule
    BF~\cite{lowe2004distinctive}&  7 \quad \quad \quad \quad  $0.449 \pm 0.007$ \\
    \textbf{SuperGlue}&  \textbf{20} \quad \quad \quad \quad  $\textbf{0.988} \pm \textbf{0.010}$ \\
    \bottomrule
  \end{tabular}
  \caption{A comparison between brute force matcher and SuperGlue. SuperGlue outperforms the brute force matcher in terms of both evaluation metrics: the average number of good matches and the average confidence score across the entire test dataset. }
  \label{tab:results}
\end{table}

\section{Conclusion}
\label{sec:print}

To sum up, our radiomic keypoint detection algorithm provides a solution for automated keypoint detection in MRI scans, overcoming challenges encountered by traditional and other deep learning methods. The limited set of radiomic keypoints facilitates training SuperRetina for increased keypoint detection. These keypoints are consistent and deformation resilient. 

Our approach paves the way for the application of GNN-based models on brain images, offering a faster and more parameter-efficient alternative compared to CNNs and transformers. Moreover, the detection of keypoints contributes to various tasks, including registration as justified in this work.

\vfill
\pagebreak

\bibliographystyle{IEEEbib}
\bibliography{strings,refs}

\end{document}